\documentclass[vecphys]{svmult}
\usepackage{graphicx}

\begin{document}

\title*{The Context of the Local Volume:\\
        Structures and Motions in the Nearby Universe} 
\titlerunning{The Context of the Local Volume} 
\author{Matthew Colless}
\institute{Anglo-Australian Observatory, PO Box 296, Epping, NSW 1710,
  Australia \\ \texttt{colless@aao.gov.au}}

\maketitle

\section{The 6dFGS and 2MRS redshift surveys}
\label{sec:1}

\begin{figure}
\centering
\includegraphics[width=\textwidth]{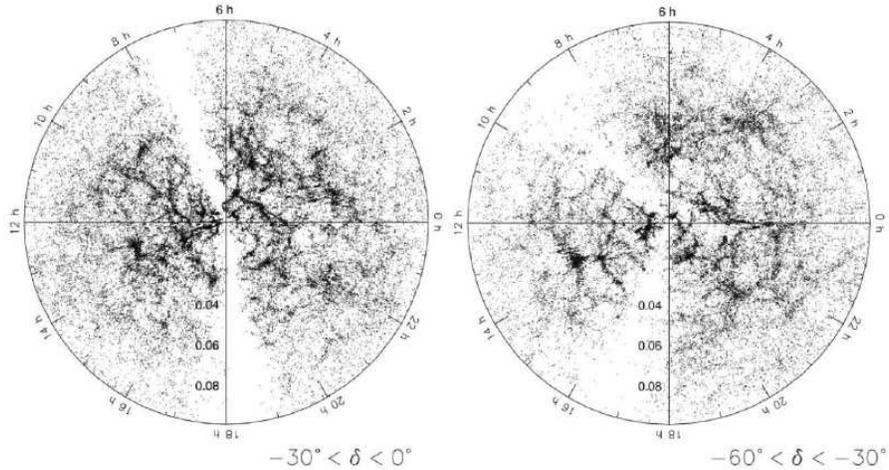}
\caption{Two slices through the 6dFGS volume showing the structures out
  to $z$=0.1 in the declinations ranges $-30^\circ$ to $0^\circ$ and
  $-60^\circ$ to $-30^\circ$. The empty sectors are the projection of
  the $|b|<10^\circ$ Zone of Avoidance around the Galactic Plane.}
\label{fig:zslice}
\end{figure}

The 6dF Galaxy Survey (6dFGS; www.aao.gov.au/local/www/6df; Jones
et~al.\ 2004, 2005) is a redshift and peculiar velocity survey of the
local universe. The observations were obtained during 2001--2006 using
the AAO's UK Schmidt Telescope and the 6dF spectrograph (Watson et~al.\
1998). The 6dFGS covers 92\% of the southern sky with
$|b|$$>$$10^\circ$. Its primary sample is from the 2MASS Extended Source
Catalog (XSC; Jarrett et~al.\ 2000) and consists of galaxies with
$K_{\rm tot}$$<$12.65. For this sample the redshift completeness is 88\%
and the median redshift is $z\approx0.05$. The 6dFGS also includes
secondary samples to comparable limits of $H$$<$12.95, $J$$<$13.75 (from
the 2MASS XSC) and $r_F$$<$15.60, $b_J$$<$16.75 (from the SuperCosmos
Sky Survey; Hambly et~al.\ 2001). The 6dFGS peculiar velocity survey is
using the Fundamental Plane to derive distances and velocities for about
15,000 bright early-type galaxies. The 6dFGS database comprises
approximately 137,000 spectra and 124,000 galaxy redshifts, plus
photometry and images. The final data release will be made public in
August 2007 (Jones et~al.\ in prep.; see www-wfau.roe.ac.uk/6dfgs).
Figure\,\ref{fig:zslice} shows the fine detail of the large-scale
structure that is revealed by two slices through the local volume
surveyed by the 6dFGS.

The 2MASS Redshift Survey (2MRS; Erdo\v{g}du et~al.\ 2006a,b) is an
all-sky survey that so far has obtained redshifts for 23,150 of the
24,773 2MASS galaxies with extinction-corrected magnitudes brighter than
$K_S$=11.25 (i.e.\ 1.4\,mag brighter than 6dFGS). Almost all of the 1600
galaxies without redshifts are at low Galactic latitudes ($|b|<5^\circ$)
or are obscured/confused by the dust and high stellar density towards
the Galactic Centre. Northern galaxies are being observed by the Whipple
Observatory 1.5m telescope, the Arecibo 305m telescope and the Green
Bank 100m telescope; in the south, most galaxies at high Galactic
latitude (about 6000 galaxies with $|b|>10^\circ$) were observed as a
part of the 6dFGS, while those at low Galactic latitudes are being
observed at CTIO. The 2MRS has a median redshift of $z\approx0.02$
(6000\,km\,s$^{-1}$) and is the densest all-sky redshift survey to date.

\section{The density field in the local universe}

Erdo\v{g}du et~al.\ (2006a,b) have used the 2MRS to reconstruct the
density field in the local volume and to predict the corresponding
velocity field and the dipole of the Local Group motion. The
reconstruction is based on the linear theory of structure formation
using a technique following that of Fisher et~al.\ (1995). The 3D
density field in redshift space is decomposed into spherical harmonics
and Bessel functions, and then the real-space density field is
constructed using a Wiener filter. This combination of spherical
harmonics and Wiener filtering smooths the noisy data and gives an
optimal reconstruction in the sense of minimising the variance between
the true and reconstructed fields. It also allows the treatment of
linear redshift distortions and greatly reduces the statistical
uncertainties and errors introduced by non-linear effects. The 2MRS is
well-suited to this approach due to its near all-sky coverage; the
unsurveyed regions close to the Galactic Plane are masked by
interpolating the distribution of neighbouring galaxies. The galaxy
power spectrum assumed for the reconstruction has a CDM shape with
$\Gamma$=0.2, $\sigma_8$=0.7 and $\beta$=0.5; comparison of the
reconstruction to the data gives a reduced $\chi^2$ of 1.3.

\begin{figure}
\centering
\includegraphics[width=0.9\textwidth]{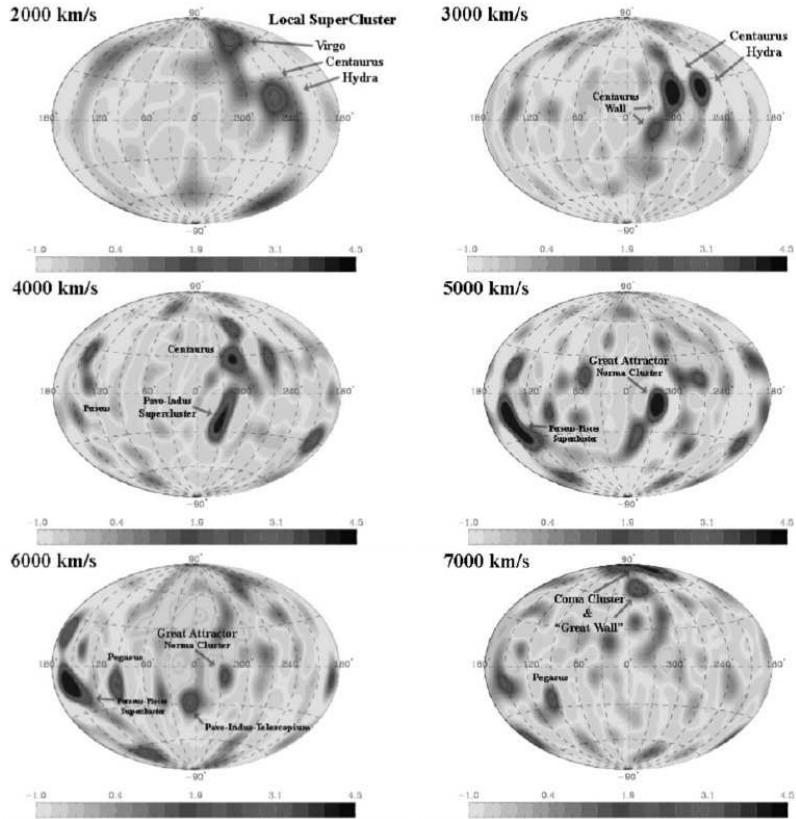}
\caption{The Wiener-filtered density field of the local volume in
  1000\,km\,s$^{-1}$ thick redshift shells over the range
  $cz$=2000--7000\,km\,s$^{-1}$, from the Virgo cluster out to the Coma
  cluster (Erdo\v{g}du et~al.\ 2006b).}
\label{fig:denshells}
\end{figure}

Figure\,\ref{fig:denshells} shows the reconstructed real-space density
field in redshift shells of thickness 1000\,km\,s$^{-1}$ over the range
$cz$=2000--7000\,km\,s$^{-1}$. The main structures in this volume are
labelled in Figure\,\ref{fig:denshells}, and include the nearby Virgo,
Hydra and Centaurus clusters, the Great Attractor region around the
Norma cluster, the Perseus-Pisces and Pavo-Indus-Telescopium
superclusters, and, in the outermost shell, the Coma cluster and the
Great Wall. Also of interest are the large low-density void regions, in
particular the network of voids including the Local Void that dominates
the lowest-redshift shell. Detailed discussion of these structures can
be found in Erdo\v{g}du et~al.\ (2006b).

\section{The peculiar velocity field}

\begin{figure}
\centering
\includegraphics[width=\textwidth]{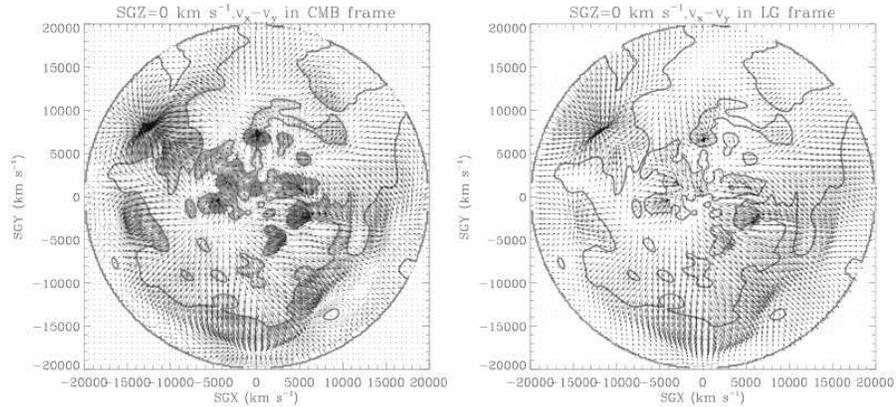}
\caption{The predicted velocity field in the Supergalactic Plane out to
  a distance of 20,000\,km\,s$^{-1}$, shown in the restframes of (at
  left) the CMB and (at right) the Local Group (Erdo\v{g}du et~al.\
  2006b).}
\label{fig:velslice}
\end{figure}

The peculiar velocity field emerges simply and naturally from the Fisher
et~al.\ (1995) method, since the harmonic expansion of the radial
velocity field can be derived from that of the gravity field, which in
turn, under linear theory assumptions, is proportional to the harmonic
expansion of the density field.

Figure\,\ref{fig:velslice} shows the predicted velocity field obtained
from the 2MRS in the Supergalactic Plane out to
$cz$=20,000\,km\,s$^{-1}$; velocities are shown in both CMB and Local
Group (LG) frames. The choice of reference frame is significant: at
smaller distances it is easier to see the real nature of the structures
in the CMB frame than in the LG frame, as the LG velocity dominates; at
larger distances, galaxies have positive CMB radial velocities, so it is
easier to compare the velocities in the LG frame than in the CMB frame.
There is a clear prediction for backside infall into the Great Attractor
region (this is most clearly visible in the LG frame).

Together, the 2MRS and 6dFGS provide the best and most detailed
information on the structures in the local universe to date. Future work
will focus on comparing the predicted velocity field from the 2MRS with
the observed velocity field from the 6dFGS.

\subsection*{Acknowledgements} 

I thank the 6dFGS team (www.aao.gov.au/local/www/6df/Team) and the 2MRS
team (cfa-www.harvard.edu/$\sim$huchra/2mass/team.php) for allowing me
to present results from both projects.


\begin{thebibliography}{99.}

\bibitem{journal} P. Erdo\v{g}du, J.P. Huchra, O. Lahav, M. Colless et~al:
     MNRAS \textbf{368}, 1515 (2006a)

\bibitem{journal} P. Erdo\v{g}du, O. Lahav, J.P. Huchra, M. Colless et~al:
     MNRAS \textbf{373}, 45 (2006b)

\bibitem{journal} K.B. Fisher, O. Lahav, Y. Hoffman, D. Lynden-Bell, S.
  Zaroubi: MNRAS \textbf{272}, 885 (1995)

\bibitem{journal} N.C. Hambly, H.T. MacGillivray, M.A. Read et~al:
     MNRAS \textbf{326}, 1295 (2001)

\bibitem{journal}  T.H. Jarrett, T. Chester, R. Cutri et~al:
     AJ \textbf{119}, 2498 (2000)

\bibitem{journal} D.H. Jones, W. Saunders, M. Colless et~al:
     MNRAS \textbf{355}, 747 (2004)

\bibitem{journal} D.H. Jones, W. Saunders, M.A. Read, M. Colless:
     PASA \textbf{22}, 277 (2005)

\bibitem{journal} F.G. Watson, Q.A. Parker, S. Miziarski:
     SPIE \textbf{3355}, 834 (1998)

\end{thebibliography}
\end{document}